# Observation of Collective-Emission-Induced Cooling inside an Optical Cavity

Hilton W. Chan, Adam T. Black, and Vladan Vuletić

*Department of Physics, Stanford University, Stanford, California 94305-4060*

(Dated: August 14, 2002)

## Abstract

We report the observation of collective-emission-induced, velocity-dependent light forces. One third of a falling sample containing $3 \times 10^6$ cesium atoms illuminated by a horizontal standing wave is stopped by cooperatively emitting light into a vertically oriented confocal resonator. We observe decelerations up to 1500 m/s$^2$ and cooling to temperatures as low as 7 $\mu$K, well below the free space Doppler limit. The measured forces substantially exceed those predicted for a single two-level atom.

In conventional free-space Doppler cooling [1], a two-level atom irradiated with laser light tuned slightly below the atomic transition frequency preferentially absorbs photons from the beam opposing the atom's velocity. The associated momentum transfer from the incident light onto the atom results in a velocity-dependent absorptive force for the atom's center-of-mass motion.

For atoms inside a resonator, the frequency variation of the electromagnetic mode density, and consequently of the atomic emission rate [2–5], can give rise to emission-induced forces [6–13], as observed for a single atom in a high-finesse resonator [10]. In the classical limit of low atomic-transition saturation, cavity Doppler cooling may occur [8, 11]: A dissipative force arises from the two-photon momentum transfer in coherent scattering, i.e., from the combined absorption and re-emission process. The atom will be cooled if the cavity is blue detuned relative to the incident light by the two-photon Doppler shift, thereby enhancing the emission of high-energy photons [12]. The light-atom detuning and the atomic structure, on the other hand, determine only the scattering rate and hence the cooling force magnitude [11].

More generally, cooling will occur whenever the average emitted light frequency exceeds the incident frequency. An interesting situation arises in the presence of intracavity gain provided by a many-atom system, which can lead to collective emission into the resonator. The optical gain amplifies the resonator-induced force $f$, while the reduced bandwidth [14] increases the velocity dependence $\partial f/\partial v$ via stronger discrimination between the red and blue Doppler sidebands. Both of these features can be expected to improve the cooling performance [15].

In this Letter, we report on the first observation of collective-emission-induced, velocity-dependent forces acting on atoms inside a resonator. The slowing of $10^6$ cesium atoms is accompanied by cooperative emission into a near-confocal resonator. A deceleration of 1500 m/s$^2$ is measured along the resonator axis and perpendicular to the incident light. We also observe resonator-induced cooling of the slowed atoms. Neither the collective emission nor the observed temperatures down to 7 $\mu$K can be explained by a single-atom model of cavity Doppler cooling [8, 11] that, for our parameters, predicts decelerations below 100 m/s$^2$ and temperatures near 200 $\mu$K [12]. Our observations suggest that the cavity-induced force is substantially enhanced by stimulated emission.

At the heart of the experiment (Fig. 1) is a 7.5 cm long, vertically oriented, nearly confocal optical resonator. Its $TEM_{00}$ mode has a waist size of $w_0 = 101$ $\mu$m and a single-mode linewidth of $\kappa/2\pi = 2$ MHz, corresponding to a finesse $F = 1000$. One cavity mirror is mounted on a piezoelectric tube providing more than one free spectral range of cavity tuning. From the observed spatial transmission pattern as a function of cavity length [16], we estimate our cavity to be $\epsilon_x =$ -24 $\mu$m and $\epsilon_y =$ -28 $\mu$m short of confocality in the $x$ and $y$ directions, respectively, where the small difference is probably due to stress-induced change in mirror curvature. Mirror spherical aberration leads to a quadratic dependence of the mode frequency on transverse mode number [17]. In combination with the deviations $\epsilon_x, \epsilon_y$ from confocality, this broadens the cavity spectrum to about 200 MHz, as measured with a 1 cm large collimated input beam. The mode density is maximized at a detuning of -200 MHz relative to the $TEM_{00}$ mode, where the aberration-limited cooling volume [5, 16] extends 2.5 mm $\times$ 800 $\mu$m $\times$ 7.5 cm in the $x$, $y$ and $z$ directions, respectively.

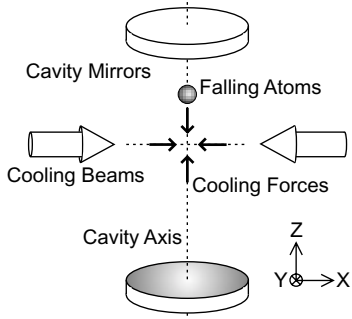

FIG. 1: Experimental setup for observing emission-induced forces on atoms. Cs atoms, illuminated by two horizontal laser beams and falling along the vertical resonator axis, experience velocity-dependent forces in the $xz$-plane.

The incident cooling beams are derived from a tunable distributed Bragg-reflector diode laser (SDL-5714-H1 from JDS Uniphase), whose linewidth is narrowed via optical feedback [18] to less than 10 kHz. Its frequency is actively stabilized relative to a Cs atomic transition. A linearly polarized standing wave, formed by a retroreflected, horizontal beam with 600 $\mu$m waist size and single-beam power of up to 16 mW, intersects the cavity axis near the cavity center. A small fraction of the light is coupled into the resonator to electronically lock the laser-cavity detuning $\Delta_c = \omega_i - \omega_c$, where $\omega_i$ and $\omega_c$ are the angular frequencies of the incident light and the nearest $TEM_{00}$ cavity resonance, respectively. Before each measurement cycle, the locking light is turned off while the cavity length is held constant.

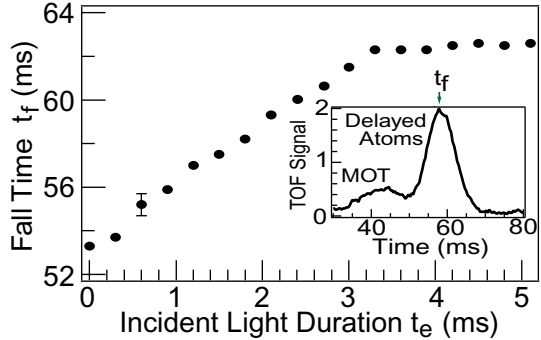

FIG. 2: Fall time $t_f$ of atoms with $v_0 = 15$ cm/s as a function of light exposure time $t_e$ for $\Delta_c/2\pi$ = -150 MHz, $I/I_s = 16$, and light-atom detuning $\delta_a/2\pi$ = - 63 MHz. For $t_e > 3.3$ ms the delayed cloud is stopped. The inset shows the TOF signal of the remnant MOT cloud and the delayed atoms for $t_e = 2$ ms.

We begin each measurement by collecting $3 \times 10^6$ Cs atoms in a magneto-optical trap (MOT), and dropping them from a variable height (0 to 5 mm) above the cavity center, where the available drop time is limited by the thermal cloud expansion that reduces the overlap with the resonant cavity modes. Upon reaching the cavity center with a velocity $v_0$ between 0 and 32 cm/s, the atoms are illuminated with light red detuned up to $\delta_a/2\pi$ = -160 MHz relative the atomic $F_g = 4 \rightarrow F_e = 5$ transition for durations between 100 $\mu$s and 25 ms. Simultaneously, the MOT repumping laser is applied on the $F_g = 3 \rightarrow F_e = 4$ transition to keep the atoms in the upper hyperfine state $F_g = 4$. Finally, we measure the atomic velocity distribution via a time-of-flight (TOF) measurement using a light sheet 2 cm below the cooling region [19], or image the atomic spatial distribution in the $xz$-plane onto a charge-coupled device camera.

The signature of a resonator-induced force on the atoms is the appearance of a second, delayed peak in the TOF signal (see inset to Fig. 2), corresponding to the slowing or stopping of a fraction of the falling cloud, that depends critically on the light-cavity detuning $\Delta_c$. For -200 MHz $< \Delta_c/2\pi < 0$ we observe a delayed TOF signal containing up to 30% (15%) of the atoms for vertical (horizontal) polarization of the incident standing wave. Tuning the cavity off resonance, we find only heating at a rate consistent with recoil heating by free-space scattering. A spatial image, taken 10 ms after the exposure, reveals two separate falling clouds. Fig. 2 shows $t_f$, the time between the extinction of the applied light and the arrival of the delayed peak to the TOF beam, as a function of light exposure time $t_e$ for $\Delta_c/2\pi = -150$ MHz, an initial velocity of $v_0 = 15$ cm/s, a light-atom detuning $\delta_a/2\pi = -63$ MHz, and a single-beam intensity $I = 16I_s$, where $I_s = 1.1$ mW/cm$^2$ is the saturation intensity. For short times $t_e$, the positive slope $\partial t_f/\partial t_e$ indicates that the atoms decelerate from $v_0$, while the constant fall time $t_f$ for large exposure times $t_e > 3.3$ ms indicates that the cloud has been at least very nearly stopped. The stopped atoms can be held in space for up to 25 ms. While we observe collective emission for both $\delta_a > 0$ and $\delta_a < 0$, we observe slowing only for an incident beam detuning between -160 MHz $< \delta_a/2\pi <$ -15 MHz and an exact retroreflection of the incident beam. For a misalignment of a few mrad, the atoms are accelerated downward or even upward (arriving up to 10 ms before the remnant MOT cloud, or up to 5 ms later than a stopped cloud). When the scattering rate into free space per atom $\Gamma_{fs}$ is increased, the observed deceleration (initial slope) is larger, and the stopping time shorter. At $\Gamma_{fs} = 3 \times 10^6$ s$^{-1}$, atoms moving at a velocity of 15 cm/s are stopped in 100 $\mu$s, indicating a vertical deceleration of 1500 m/s$^2$.

Using two calibrated photodiodes, we measure the ratio $\eta = \Gamma_c/\Gamma_{fs}$ of the emission rates into the cavity $\Gamma_c$ and into free space $\Gamma_{fs}$. Surprisingly, whenever a substantial sample fraction is delayed, we observe collective emission into the cavity, as identified by a sharp increase in $\eta$ above a certain threshold incident intensity $I_{th}$ (Fig. 3). We find that $I_{th}$ is inversely proportional to $\delta_a{}^2$, and that it corresponds to $\Gamma_{fs} = 2 \times 10^5$ s$^{-1}$ per atom for $10^6$ atoms. When varying the drop height and temperature $T_{MOT}$ of the MOT we find that $I_{th}$ is independent of the mean initial velocity $v_0$ of the falling cloud, and inversely proportional to $T_{MOT}$ in the region 7 $\mu$K $< T_{MOT} <$ 30 $\mu$K. As the atomic transition approaches saturation, the emission ratio $\eta$ slowly decreases again.

The increase in cavity emission rate $\Gamma_c = \eta\Gamma_{fs}$ implies the possibility of a corresponding enhancement over the single-atom cooling force. In our near-confocal setup with frequency-dependent electromagnetic mode density $\rho(\omega)$, the value of $\eta$ varies with laser-cavity detuning $\Delta_c$. Below the collective emission threshold, we find a maximum $\eta_s \approx 0.05$ near $\Delta_c =$ -200 MHz, consistent with a value derived from the measured $\rho(\omega)$ and $TEM_{00}$ linewidth $\kappa$. Above threshold, we observe an increase by a factor of 20 up to $\eta_c \approx 1$. For a single atom moving at $v_0 = 15$ cm/s, the expected deceleration [12] is $a_s = 2\eta_s\Gamma_{fs}v_{rec}kv_0/\kappa = 90$ m/s$^2$ at $\Gamma_{fs} = 3 \times 10^6$ s$^{-1}$, where $v_{rec} = 3.5$ mm/s is the recoil velocity, and $2\pi/k = 852$ nm. The 17 times larger measured value $a = 1500$ m/s$^2$ suggests that the collective emission with $\eta_c/\eta_s = 20$ may be responsible for the large observed force.

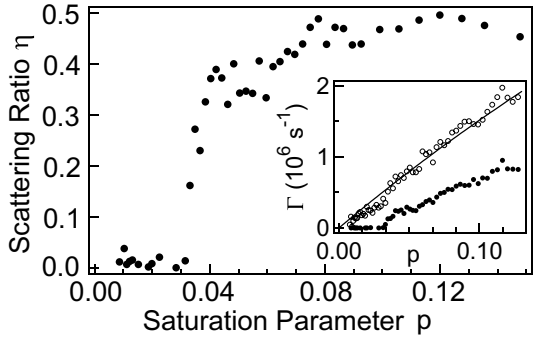

FIG. 3: Scattering ratio $\eta = \Gamma_c/\Gamma_{fs}$ versus atomic saturation parameter $p$ for $10^6$ atoms, and laser-atom and laser-cavity detunings of $\delta_a/2\pi = $ -78 MHz and $\Delta_c/2\pi = $ -150 MHz, respectively. At $p = 0.03$, collective emission increases $\eta$ abruptly. The inset shows the scattering rate into free space $\Gamma_{fs}$ (open circles) and into the cavity $\Gamma_c$ (solid circles). The solid line is the calculated $\Gamma_{fs}$.

The sudden increase in $\eta$ is consistent with Raman lasing [20] between magnetic sublevels. Due to different Clebsch-Gordan coefficient magnitudes, the linearly polarized incident light optically pumps the atoms towards states of lower magnitude $|m|$, where $m$ is the magnetic quantum number along the polarization axis. The population differences result in Raman gain for circularly polarized light on the transitions $|F_g = 4, m\rangle \to |F_e = 5, m\rangle \to |F_g = 4, m'\rangle$, where $m' = m + 1$ ($m' = m - 1$) for $m \geq 0$ ($m \leq 0$). If all magnetic sublevels were degenerate, then the emitted photon on one transition, for example $|m| \to |m| + 1$, would be reabsorbed by the $-|m| \to -|m| - 1$ transition, destroying the gain mechanism. However, the incident-beam-induced light shifts break this degeneracy, suppressing the reabsorbing two-photon transition. Several observations agree with this interpretation. First, for vertical incident polarization along $z$, the emitted circularly polarized light couples more strongly to the resonator than for horizontal incident polarization, leading to stronger lasing and a larger number of slowed atoms. Second, for $z$ polarization a magnetic field of 0.4 G applied in the transverse $xy$-plane inhibits laser action by causing Larmor precession that destroys the population inversion, while similar or larger fields applied along the $z$ direction enhance the laser emission and slowing force. Furthermore, using microwave transitions $|F_g = 4, m\rangle \to |F_g = 3, m\rangle$ in combination with fluorescent detection on the $F_g = 3 \to F_e = 2$ transition, we found the population ratio of $|4, 2\rangle$ to $|4, 1\rangle$ strongly increases as the lasing threshold is crossed. Finally, the observed cessation of collective emission at large red detuning $\delta_a/2\pi < $ -160 MHz is simply explained by depolarization of the atomic sample

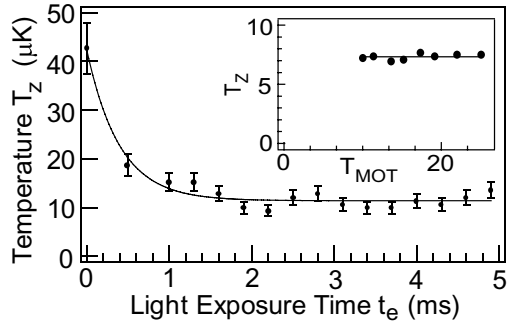

FIG. 4: Evolution of the delayed atom sample's vertical temperature $T_z$ for $\Delta_c/2\pi =$ -150 MHz, $I/I_s = 20$, and $\delta_a/2\pi =$ -78 MHz. The temperature decreases with a time constant $\tau = 0.4$ ms. The inset shows $T_z$ versus initial MOT temperature $T_{MOT}$ in $\mu$K for $I/I_s = 13$ and $\delta_a/2\pi =$ -58 MHz with all other parameters as given in Fig. 5.

due to excitation to the $F_e = 4$ hyperfine excited state and decay to $F_g = 3$.

When we vary the light exposure time we observe not only a slowing of the cloud (Fig. 2), but also a changing width of the approximately Gaussian delayed peak (see inset to Fig. 2). We use the rms velocity to assign a temperature, and plot in Fig. 4 the temperature evolution for $\Delta_c/2\pi =$ -150 MHz, $I/I_s = 20$, and a light-atom detuning $\delta_a/2\pi =$ -78 MHz. During the evolution, the number of delayed atoms (measured via the area of the peak) does not change by more than 50 %. The time constant $\tau = 0.4$ ms is about 100 times longer than the scattering time. We typically observe vertical temperatures of $T_z = 16$ $\mu$K for instantaneous cooling light extinction, and down to $T_z = 7$ $\mu$K for slow cooling light extinction with a time constant longer than 0.4 ms. Cooling to similar final temperatures has been observed for MOTs with initial temperatures as high as 63 $\mu$K. By spatially imaging the falling atoms, we confirm cooling along the incident light direction as predicted by cavity Doppler cooling [12]. For example, for a MOT with initial temperature 46 $\mu$K (51 $\mu$K) in the $x$ direction ($z$ direction) and instantaneous extinction of incident light, we measure a final temperature of 20 $\mu$K (14 $\mu$K). In contrast to polarization gradient cooling, the final temperature is largely independent of cooling beam intensity and light-atom detuning $\delta_a$ for -160 MHz $< \delta_a/2\pi <$ -50 MHz (Fig. 5), and only increases as the atomic transition approaches saturation.

Although the final temperatures are not substantially below the initial MOT temperature, for long exposure times the heating due to free-space scattering is substantial, as observed with the resonator tuned off-resonance. For instance, for an off-resonant cavity the final temperature after 5 ms for the parameters of Fig. 4 is 125 $\mu$K. Therefore the observed temperature and approximately constant atom number cannot be explained by filtering from the MOT distribution and trapping in the intracavity standing wave.

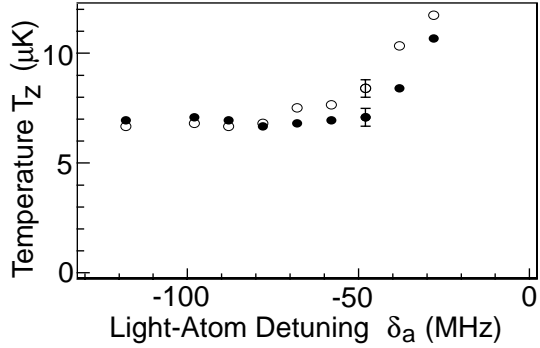

FIG. 5: Vertical temperature $T_z$ versus light-atom detuning $\delta_a$ for $I/I_s = 22$ (open circles) and $I/I_s = 11$ (solid circles) along with typical error bars. All data were taken with $10^6$ atoms for $\Delta_c/2\pi = $ -200 MHz, 5 ms exposure time, and slow light extinction with $\tau = 0.4$ ms.

The single-atom model of cavity cooling [8] predicts a friction force only in a region of steep positive cavity slope $\partial\eta/\partial\omega$, where the scattering into the cavity increases rapidly with emission frequency $\omega$ [11]. In contrast, we observe slowing and cooling if the incident light frequency lies anywhere within the 200 MHz wide multimode line shape of our near-confocal resonator. The number of atoms cooled, however, is half as large in the region with $\partial\eta/\partial\omega < 0$, where the single-atom model predicts heating. Furthermore, the observed temperature is 20 times lower than the expected final value [12] of $T_{s,z} = \hbar\kappa/(10k_B\eta_{nl}) \approx$ 190 $\mu$K. For an off-resonant cavity, only heating of the falling cloud is observed, indicating that the incident light does not decelerate, trap or cool the atoms independently of the resonator. Conventional cooling mechanisms cannot produce any strong dissipative forces in the vertical direction, perpendicular to the applied laser beams. Free-space Doppler cooling in our standing wave at $\delta_a/2\pi = $ -160 MHz should lead to a horizontal temperature [19] of 3.8 mK, more than 300 times higher than the observed value. Lower $x$ temperatures could be achieved by sub-Doppler cooling [19, 21], but the incident light contains no polarization gradients. Due to the incident-beam-induced differential light shift $\Delta U/h \approx 1$ MHz between neighboring magnetic sublevels, any polarization interference pattern between the Raman emission into the resonator and the incident light is not stationary, and should not lead to conventional polarization gradient cooling. However, a novel form of sub-Doppler cooling involving time-varying polarization gradients acting on non-degenerate magnetic sublevels cannot be excluded.

In conclusion, we have observed collective emission from cold atoms inside an optical resonator, accompanied by velocity-dependent, emission-induced forces that are significantly stronger than expected for single atoms. A model for the observed resonator-induced slowing and cooling must account for both the weak dependence on cavity slope and the requirement of negative light-atom detuning. Possible explanations include self-organization of the atoms into patterns that maximize collective scattering [22], or the joint effects of cavity tuning by the moving atoms [8] and coherent Raman scattering, possibly in combination with laser mode competition between the blue and red Doppler emission sidebands.

This work was supported in parts by the ARO under MURI grant Y-00-0005-02. We would like to thank Xinan Wu for technical assistance and Cheng Chin for stimulating discussions.

---